\documentstyle[12pt,epsbox]{article}

\catcode`\@=11
\long\def\@makefntext#1{
\protect\noindent \hbox to 3.2pt {\hskip-.9pt
$^{{\ninerm\@thefnmark}}$\hfil}#1\hfill}                

\def\@makefnmark{\hbox to 0pt{$^{\@thefnmark}$\hss}}  

\def\ps@myheadings{\let\@mkboth\@gobbletwo
\def\@oddhead{\hbox{}
\rightmark\hfil\ninerm\thepage}
\def\@oddfoot{}\def\@evenhead{\ninerm\thepage\hfil
\leftmark\hbox{}}\def\@evenfoot{}
\def\sectionmark##1{}\def\subsectionmark##1{}}

\renewcommand{\thefootnote}{\fnsymbol{footnote}}

\newcounter{sectionc}\newcounter{subsectionc}\newcounter{subsubsectionc}
\renewcommand{\section}[1] {\vspace*{0.6cm}\addtocounter{sectionc}{1}
\setcounter{subsectionc}{0}\setcounter{subsubsectionc}{0}\noindent
        {\normalsize\bf\thesectionc. #1}\par\vspace*{0.4cm}}
\renewcommand{\subsection}[1] {\vspace*{0.6cm}\addtocounter{subsectionc}{1}
        \setcounter{subsubsectionc}{0}\noindent
        {\normalsize\it\thesectionc.\thesubsectionc. #1}\par\vspace*{0.4cm}}
\renewcommand{\subsubsection}[1]
{\vspace*{0.6cm}\addtocounter{subsubsectionc}{1}
        \noindent
{\normalsize\rm\thesectionc.\thesubsectionc.\thesubsubsectionc
{}.
        #1}\par\vspace*{0.4cm}}

\newcounter{appendixc}
\newcounter{subappendixc}[appendixc]
\newcounter{subsubappendixc}[subappendixc]

\renewcommand{\appendix}[1] {\vspace*{0.6cm}
        \refstepcounter{appendixc}
        \setcounter{figure}{0}
        \setcounter{table}{0}
        \setcounter{equation}{0}
        \renewcommand{\thefigure}{\Alph{appendixc}.\arabic{figure}}
        \renewcommand{\thetable}{\Alph{appendixc}.\arabic{table}}
        \renewcommand{\theappendixc}{\Alph{appendixc}}
        \renewcommand{\theequation}{\Alph{appendixc}.\arabic{equation}}
        \noindent{\bf Appendix \theappendixc #1}\par\vspace*{0.4cm}}

\def\abstracts#1{{
        \centering{\begin{minipage}{12.2truecm}
        \footnotesize\baselineskip=12pt\noindent
        \centerline{\footnotesize ABSTRACT}\vspace*{0.3cm}
        \parindent=0pt #1
        \end{minipage}}\par}}

\renewenvironment{thebibliography}[1]
        {\begin{list}{\arabic{enumi}.}
        {\usecounter{enumi}\setlength{\parsep}{0pt}
\setlength{\leftmargin 1.25cm}{\rightmargin 0pt}
         \setlength{\itemsep}{0pt} \settowidth
        {\labelwidth}{#1.}\sloppy}}{\end{list}}

\newcounter{itemlistc}
\newcounter{romanlistc}
\newcounter{alphlistc}
\newcounter{arabiclistc}

\newcommand{\fcaption}[1]{
        \refstepcounter{figure}
        \setbox\@tempboxa = \hbox{\footnotesize Fig.~\thefigure. #1}
        \ifdim \wd\@tempboxa > 6in
           {\begin{center}
        \parbox{6in}{\footnotesize\baselineskip=12pt Fig.~\thefigure. #1}
            \end{center}}
        \else
             {\begin{center}
             {\footnotesize Fig.~\thefigure. #1}
              \end{center}}
        \fi}

\newcommand{\tcaption}[1]{
        \refstepcounter{table}
        \setbox\@tempboxa = \hbox{\footnotesize Table~\thetable. #1}
        \ifdim \wd\@tempboxa > 6in
           {\begin{center}
        \parbox{6in}{\footnotesize\baselineskip=12pt Table~\thetable. #1}
            \end{center}}
        \else
             {\begin{center}
             {\footnotesize Table~\thetable. #1}
              \end{center}}
        \fi}

\def\@citex[#1]#2{\if@filesw\immediate\write\@auxout
        {\string\citation{#2}}\fi
\def\@citea{}\@cite{\@for\@citeb:=#2\do
        {\@citea\def\@citea{,}\@ifundefined
        {b@\@citeb}{{\bf ?}\@warning
        {Citation `\@citeb' on page \thepage \space undefined}}
        {\csname b@\@citeb\endcsname}}}{#1}}

\newif\if@cghi
\def\cite{\@cghitrue\@ifnextchar [{\@tempswatrue
        \@citex}{\@tempswafalse\@citex[]}}
\def\citelow{\@cghifalse\@ifnextchar [{\@tempswatrue
        \@citex}{\@tempswafalse\@citex[]}}
\def\@cite#1#2{{$\null^{#1}$\if@tempswa\typeout
        {IJCGA warning: optional citation argument
        ignored: `#2'} \fi}}

 1
 1
 1

\font\ninerm=cmr9

\begin{document}

\begin{flushleft}
\large
{SAGA-HE-82-95
\hfill June 20, 1995}  \\
\end{flushleft}

\vspace{3.0cm}

\begin{center}

\LARGE{{\bf Nuclear modification}} \\

\vspace{0.3cm}
\LARGE{{\bf in the structure function $\bf F_3$}} \\

\vspace{2.0cm}

\Large
{S. Kumano and M. Miyama $^\star$ }   \\

\vspace{1.0cm}

\Large
{Department of Physics}         \\

\vspace{0.1cm}

\Large
{Saga University}      \\

\vspace{0.1cm}

\Large
{Saga 840, Japan} \\

\vspace{2.2cm}

\large
{Talk given at the IV International Symposium on} \\

\vspace{0.3cm}

{``Weak and Electromagnetic Interactions in Nuclei"} \\

\vspace{0.7cm}

{Osaka, Japan,  June 12 -- 16, 1995}  \\

\end{center}

\vspace{1.3cm}
\vfill

\noindent
{\rule{6.cm}{0.1mm}} \\

\vspace{-0.4cm}

\noindent
\vspace{-0.2cm}
\normalsize
{$\star$ Email: kumanos or 94sm10@cc.saga-u.ac.jp}  \\

\vspace{-0.2cm}
\noindent
\normalsize
{Our group activities are listed
 at ftp://ftp.cc.saga-u.ac.jp/pub/paper/riko/quantum1} \\
\vspace{-0.6cm}

\noindent
{or at
http://www.cc.saga-u.ac.jp/saga-u/riko/physics/quantum1/structure.html.} \\

\vspace{+0.2cm}
\hfill
{to be published in proceedings}

\vfill\eject
\centerline{\normalsize\bf NUCLEAR MODIFICATION}
\baselineskip=16pt
\centerline{\normalsize\bf IN THE STRUCTURE FUNCTION $\bf F_3$}

\centerline{\footnotesize S. Kumano and M. Miyama
\footnote{(URL)
http://www.cc.saga-u.ac.jp/saga-u/riko/physics/quantum1/structure.html, \\
(Anonymous FTP) ftp://ftp.cc.saga-u.ac.jp/pub/paper/riko/quantum1.}}
\baselineskip=13pt
\centerline{\footnotesize\it Department of Physics, Saga University}
\baselineskip=12pt
\centerline{\footnotesize\it Saga 840, Japan}
\centerline{\footnotesize E-mail: kumanos and 94sm10@cc.saga-u.ac.jp}

\vspace*{0.9cm}
\abstracts{
Nuclear shadowing in the structure function $F_3$ is investigated
as a possible test of shadowing models.
The $F_3$ shadowing is studied
in two different theoretical models:
a parton-recombination model with $Q^2$ rescaling
and an aligned-jet model.
We find that predictions in these models differ completely
in the small $x$ region.
It suggests that the $F_3$ shadowing could be used
for discriminating among different models,
which produce similar shadowing behavior
in the structure function $F_2$.
}

\normalsize\baselineskip=15pt
\setcounter{footnote}{0}
\renewcommand{\thefootnote}{\alph{footnote}}
\section{Introduction}

This talk is based on our investigation with R. Kobayashi
in Ref. 1.
Nuclear shadowing in the structure function $F_2$ has been
studied extensively last several years.\cite{SKF2}
There are various models in explaining the shadowing.
These models include a traditional explanation in terms of
vector-meson dominance and a new approach using parton
interactions in nuclei.
These viewpoints seem to be very different.
The former model indicates that the virtual photon
transforms into vector-meson states, which then interact
with a target nucleus. Because the mesons interact predominantly
in the nuclear surface, internal constituents are shadowed.
On the other hand, the latter model is based
on a parton picture.
At small $x$, the localization
size of a parton could exceed the average nucleon separation in a nucleus.
Therefore, partons in different nucleons could interact,
and these interactions are called parton recombinations.
The recombinations are extra nuclear effects,
which give rise to the nuclear shadowing.

These different models produce similar $x$ dependence
in the structure function $F_2$, so that we cannot distinguish
among the models in comparison with experimental data.
Various shadowing models may be tested by
other quantities such as sea-quark and gluon distributions
in nuclei.\cite{SKGLUE}
In this paper, we propose that the structure function $F_3$ could
be useful in determining the appropriate shadowing description.
Namely, valence-quark distributions at small $x$ are not
well investigated and they could be used for determining the shadowing model.
In order to find the possibility,
two different models are employed for the shadowing. The first one
is a parton model with parton recombination and $Q^2$ rescaling
effects,\cite{SKF2} and the other is an aligned-jet model \cite{AJMF3}
which is an extension of the vector-meson-dominance model.
Nuclear modification
of $F_3(x)$ in these models is explained in section 2.1.
Then, we discuss whether
or not the results are compatible with current experimental data.
Experimental restrictions are estimated in section 2.2 and the results
are compared with the theoretical predictions.

\section{Shadowing in the structure function $\bf F_3$}

Nuclear modification of $F_3$ could be estimated from
$F_2$ at medium and large $x$ because valence-quark
distributions dominate both structure functions.
Hence, the essential point of our study
is to investigate $F_3$ in the small $x$ region.
We neglect next-to-leading order corrections in $F_3$ for simplicity,
so that $F_3$ is given by the valence-quark distributions:
$F_3(x)=u_v(x)+d_v(x)$.
Therefore, shadowing in $F_3$ indicates
nuclear modification of the valence-quark distributions at small $x$.
We explain theoretical and experimental situations in the following.

\subsection{Model predictions}

Because there is little theoretical study on the $F_3$ shadowing,
existing theoretical predictions are limited at this stage.
Among the models, we employ two very different ones for studying $F_3$.

The first one is a hybrid parton model with $Q^2$ rescaling
and parton recombination effects.\cite{SKF2}
The rescaling model was originally proposed as a model
for explaining the medium $x$ region, and the recombination
as a model in the small $x$ region.
Combining these mechanism, we could explain nuclear structure
functions $F_2^A$ from small $x$ ($\approx 0.01$)
to large $x$ ($\approx 0.8$).
This unified model can be applied to the structure function $F_3$.
According to the $Q^2$ rescaling model,
nuclear structure functions
$F_3^A(x,Q^2)$ are given by rescaling (increasing) $Q^2$
in the nucleon structure function $F_3^N(x,Q^2)$.
Therefore, the ratio $R_3\equiv F_3^A(x)/F_3^D(x)$ is smaller
than unity at medium $x$ as it explains the EMC effect in this region.
Since the rescaling satisfies the baryon-number conservation
$\displaystyle{\int dx [u_v(x)+d_v(x)]=3}$,
the ratio $R_3$ becomes larger than unity at small $x$.
Parton-recombination contributions to $F_3(x)$ are rather contrary to
those in the $Q^2$ rescaling model in the sense that
the recombinations decrease the ratio at small $x$ and
increase it at medium and large $x$.
The overall effects are shown by a solid line (model 1) in Fig. 1.
It is interesting to note that the model predicts antishadowing
at small $x$ instead of shadowing.
In this parton model, the $F_3$ shadowing differs
distinctively from the $F_2$ one:
$F_3^A(x) / F_3^N(x) \ne F_2^A(x) / F_2^N(x)$
at small $x$.
In other words, valence-quark modification is different from
the sea-quark one.
However, there is a model which predicts $F_3$ shadowing
similar to the $F_2$ one.
An example is discussed in the following.

The second model is the aligned jet model in Ref. 4.
This model is based on the vector-meson dominance model,
which indicates that
the virtual photon transforms into vector meson states
and they interact with the target.
The propagation length of the hadronic fluctuation could exceed
the average nucleon separation in a nucleus at small $x$,
and shadowing occurs due to multiple scatterings.
In the aligned-jet model,
the virtual photon transforms into a $q\bar q$ pair,
which then interacts with the target.
However, the only $q\bar q$ pair aligned in the direction
of $\gamma$ (W) interacts in a similar way
to the vector-meson interactions with the target.
The model predication for the $F_3$ shadowing is shown
by a solid curve (model 2) in Fig. 1.
This shadowing is very similar to the $F_2$ shadowing:
$F_3^A(x) / F_3^N(x) \approx F_2^A(x) / F_2^N(x)$
at small $x$.
It is because the $q\bar q$ pair interacts with sea and valence
quarks in the similar way.
The model curve is obtained by the aligned-jet-model
together with the baryon-number conservation.
Their study to combine the shadowing mechanism with
the medium $x$ physics, such as the nuclear binding,
is still in progress.

As far as we know, these are only two papers which
discuss the $F_3$ shadowing explicitly.
Obviously, other model predictions should be investigated in future
in comparison with the above estimates.
We examine whether these model predications are compatible
with current experimental data in the following.

\subsection{Experimental restriction and comparison}

It is rather surprising that both model predications
are opposite at small $x$ even though both results are
very similar in the $F_2$ shadowing.
This fact suggests that the $F_3$ shadowing
could be used for discriminating among various models.
We can at least rule out one of the above models
by accurate experimental measurements.

\vspace{0.10cm}
\noindent
\parbox{7.5cm}
{
\hspace{0.6cm}
\baselineskip=0.212in
The structure function $F_3$ is measured in neutrino interactions.
Because the process is a weak interaction, most data are taken
by using nuclear targets with large mass number.
In order to learn about the nuclear modification,
the ratio $F_3^A(x)/F_3^D(x)$ should be taken.
Unfortunately, available deuteron data are not accurate enough
to investigate 10--20\% effects.
There is little experimental information on the $F_3$ shadowing
from neutrino data, so that
we estimate a current experimental restriction
on the $F_3$ shadowing by using
$F_2^A/F_2^D \ (\equiv R_2)$ data at medium $x$
and the baryon number conservation.
 } \ \hspace{0.5cm} \
\parbox{7.5cm}
{
\vspace{0.5cm}
\epsfile{file=fig1.eps,width=6.5cm}
\par\vspace{0.2cm}\hspace{0.8cm}
Fig. 1  ~Shadowing in $F_3(x)$.
}
\vspace{-0.1cm}

In finding the restriction, we neglect next-to-leading-order
effects and assume $R_3=R_2 [\equiv F_2^A(x)/F_2^D(x)]$
in the region $x \ge 0.3$ because
valence-quark distributions dominate both structure functions.
We employ SLAC $R_2^{Ca}$ data, which are fitted by a smooth curve.
This curve is extrapolated into the small $x$ region.
The only guideline is to satisfy the baryon-number conservation.
First, a straight line in the logarithmic $x$ is simply drawn
from $x$=0.3 as shown by the dashed line in Fig. 1
so as to satisfy the conservation.
Second, the curve is smoothly extrapolated into the small $x$
region so as to satisfy the conservation
by allowing about 6\% antishadowing at $x=0.1-0.2$.
The result is the dotted curve in Fig. 1.
The first line is roughly the upper limit of nuclear modification,
and the second curve is roughly the lower bound.
The shaded area between these curves is the area of possible
nuclear modification, which is allowed by present experimental
data of $R_2$ and the baryon-number conservation.

It is noteworthy that the first parton-model (model 1)
prediction is roughly equal to the upper bound curve,
and the aligned-jet model (model 2)
prediction is to the lower bound curve.
So the models are two extreme cases, which are both
acceptable in our present knowledge.
We have not investigated the details of other model predictions.
However, it is very encouraging to investigate
the (anti)shadowing phenomena of $F_3$ in the sense that
the observable could be useful in discriminating among
different models, which produce similar results in
the $F_2$ shadowing.

\section{Conclusions}

We have investigated nuclear modification of the structure function
$F_3$. In particular, the $F_3$ shadowing at small $x$ is studied
in two different theoretical models:
the parton-recombination with $Q^2$ rescaling and the aligned-jet model.
These models predict very different shadowing behavior, namely
antishadowing in the first model and shadowing in the second model.
Therefore, it is in principle possible to rule out one of these models
by measuring the ratio $F_3^A/F_2^D$. However, current experimental data
are not accurate enough to probe the $F_3$ shadowing.
Our investigation is merely intended to shed light on the model difference
in $F_3$ at this stage. Much detailed theoretical and experimental
investigations have to be done in the near future.

\vspace{0.4cm}
\noindent
{\bf Acknowledgment}
\vspace{0.3cm}

This research was partly supported by the Grant-in-Aid for
Scientific Research from the Japanese Ministry of Education,
Science, and Culture under the contract number 06640406.



\end{document}